\documentclass[journal]{IEEEtran}
\usepackage{amssymb}
\usepackage{textgreek}

\usepackage{adjustbox}
\usepackage{cite}
\usepackage{hyperref}
\ifCLASSINFOpdf
  \usepackage{graphicx}
  \usepackage{caption}
  
\else
\fi
\usepackage{amssymb}
\usepackage{amsmath}
\usepackage{etoolbox}
\usepackage{algorithm,algorithmicx,algpseudocode}
\usepackage[colorinlistoftodos]{todonotes}
\usepackage{multicol}
\usepackage{pgfplots}
\usepackage{tikz}
\usepackage{subcaption}
\usepackage{multirow}
\begin{document}
%\setlength{\abovedisplayskip}{0pt}
%\setlength{\belowdisplayskip}{0pt}
\title{Parameter estimation for optimal path planning in internal transportation}

\author{Pragna~Das, %~\IEEEmembership{Member,~IEEE,}
        Llu{\'\i}s~Ribas-Xirgo,~\IEEEmembership{Member,~IEEE}
        %and~Jane~Doe,~\IEEEmembership{Life~Fellow,~IEEE}% <-this % stops a space
        
\thanks{P Das is with Microelectronics and Systems Electronics Department, Autonomous University of Barcelona, Bellaterra, Barcelona, 08193 Spain e-mail: (pragna.das@uab.cat).}% <-this % stops a space
\thanks{L. Xirgo is with Microelectronics and Systems Electronics Department, Autonomous University of Barcelona, Bellaterra, Barcelona, 08193 Spain e-mail: (lluis.ribas@uab.cat).}}% <-this % stops a space
\maketitle

\begin{abstract}
The costs incurred in a mobile robot (MR) change due to change in physical and environmental factors. Usually, there are two approaches to consider these costs, either explicitly modelling these different factors to calculate the cost or consider heuristics costs. First approach is lengthy and cumbersome and requires a new model for every new factor. Heuristics cost cannot account for the change in cost due to change in state. This work proposes a new method to compute these costs, without the need of explicitly modelling the factors. The identified cost is modelled in a bi-linear state-space form where the change of costs is formed due to the change of these states. This eliminates the need to model all factors to derive the cost for every robot. In context of transportation, the travel time is identified as the key parameter to understand costs of traversing paths to carry material. The necessity to identify and estimate these travel times is proved by using them in route planning. The paths are computed constantly and average of total path costs of these paths are compared with that of paths obtained by heuristics costs. The results show that average total path costs of paths obtained through on-line estimated travel times are 15\% less that of paths obtained by heuristics costs.
\end{abstract}
%\begin{abstract}
   % Note to practioner: The motivation of this work is to identify and estimate the costs in system level where the decisions for planning takes place, though these costs arising 
%\end{abstract}[]
\IEEEpeerreviewmaketitle
\section{Introduction}
\label{intro}
\begin{figure}[h]
\centering
\includegraphics[scale = 0.36]{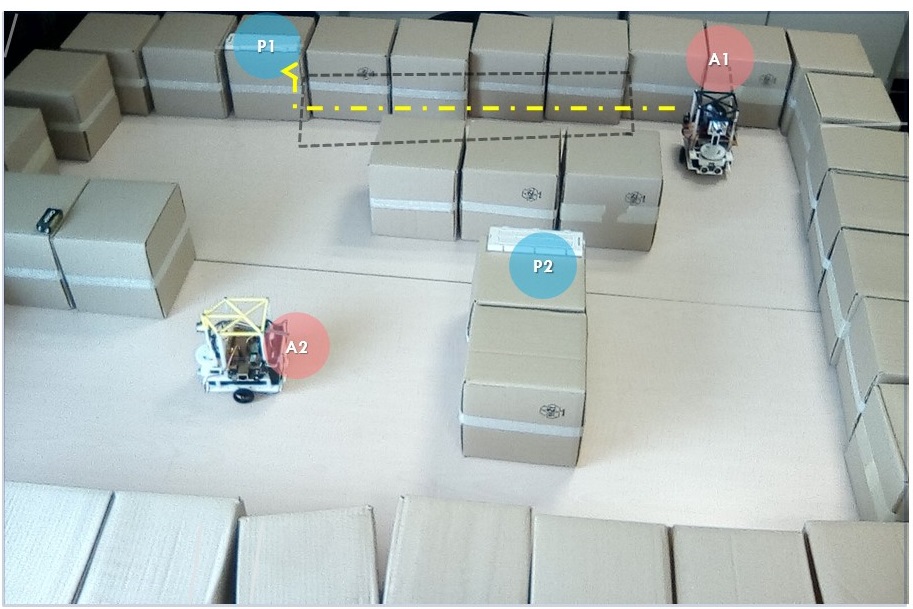}
\caption{An example}
\label{fig_exp}
\end{figure}

\IEEEPARstart{M}{obile robot} (MR) based systems used for internal logistics in factories demand cost efficient decisions on planning and co-ordination. Usually, the information about current condition of robot, floor, batteries and other robots play crucial role in decision making \cite{colby2015implicit, gerkey2004explicit, bayram2016coalition}. The costs of an MR are incurred due to performances and are influenced by these states. The conventional robot control approaches consider and derive the cost from a kinematic model \cite{eachFactorModelCost, kimCMU} in terms of battery charge ($F$1) and floor roughness ($F$2) as they are the first order factors for estimating movement costs, basically, time and energy. Usually, the state of charge of batteries can be modelled and predicted in accordance to the discharge profile \cite{aneiros2013proposed}. However, these models involve an undetermined number of coefficients which should be identified to minimise errors between measured data and estimated state of charge. Still, the discharge of batteries due to load or floor condition or similar environmental factors cannot be predicted from discharge profile. On the other hand, progressive floor conditions can be mathematically estimated but not disruptive events like oil leaks \textit{et~cetera} cannot be modelled and estimated. Thus, floor roughness, $F$2, including all factors cannot be predicted in the same way. Besides, in a traffic network, a two dimensional matrix of factors can best describe the floor condition, as in $F$2 ($i$, $j$) where $i$ and $j$ denote the source and destination spots. The final, kinematic model for a robot would be an annotated graph of cost functions $X$($i$, $j$), where $X$($i$, $j$) = $F$( $F$1, $F$2 ($i$, $j$)) that would require some parameter identification for proper cost estimation. Also, a separate model has to be devised for each new factor with this explicit modelling technique. This is potentially time consuming and cumbersome. Another conventional approach of considering cost, is obtaining heuristics cost at system level for planning, as it reduces this lengthy process. However, the heuristics cost \cite{Devaurs2015, kollmitz2015time} cannot encompass the change in these factors, as they are not estimated based on changing physical and environmental conditions. Hence, they do not represent close-to-real costs. Still, they are used in planning as they reduce the burden of explicitly modelling each physical and environmental factor. So, one way to obtain costs in robotic system is by cumbersome and time consuming kinematic modelling considering all physical and environmental factors and the other way is heuristic costs which are not representative. This work deals with directly modelling the cost to correctly estimate it rather than using heuristic cost. The proposed model does not involve kinematics of the robot but uses time as the key cost. This idea is further illustrated in the following example in Figure~\ref{fig_exp}. 
Figure~\ref{fig_exp} illustrates a scaled down automated internal transportation system executed by MRs. In this example, all MRs can execute only one task at a time. Let, at $t_i$, the path computed for $A$1 to carry some material to $P$1 is marked by the dotted line. Again, at time $t_j$ ($j>i$), $A$1 needs to carry the same material to $P$1. But now, the battery capability of $A$1 has decreased due to execution of previous tasks and the condition of the given path has deteriorated (marked by dotted rectangle). Hence, more cost in terms of time and energy will be required to reach $P$1 at $t_j$ by $A$1. The time to traverse each segment of path changed due to change in condition of battery and floor. This travel time, thus, showcases the real cost. This work focuses on suitably modelling and estimating the travel times to generate close-to-real costs, which can encompass the changes in physical and environmental factors.

\begin{figure}[h]
\centering
\includegraphics[scale = 0.5]{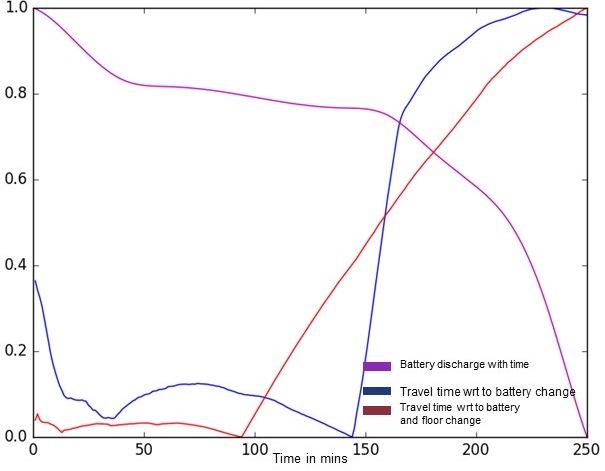}
\caption{Travel time changes with battery and floor condition}
\label{battvsTT}
\end{figure}

In an experiment conducted in our laboratory, the relation between state of charge of batteries and time taken to traverse segments of paths is observed. First, an MR was instructed to travel a particular distance repetitively till the battery is fully exhausted from complete charge. The values of the travel time was recorded in seconds every time the MR traversed that edge. This number of times is expressed as $k$. The progressive mean values of the travel time (blue curve) is plotted against the battery discharging profile (violet curve) of the Lithium-ion batteries in Figure~\ref{battvsTT}. 
Parity is observed in blue curve and violet curve where blue curve increases first with downfall of violet curve, then forms a slight hump with the violet curve settling to a steady value and then increases steadily till full discharge. In the next step, the floor condition was made rough from smooth and again the MR was instructed to travel that distance repetitively till the battery is fully exhausted. The progressive mean values of travel time (red curve), which changed with both the change of state of the charge of batteries and the floor is plotted in the same Figure~\ref{battvsTT}. The red curve also increases first with discharge at the beginning, shows steady value in between and then shoots up steadily till full discharge. The longer increase of values of travel time in red curve can be attributed to the rough floor, because at equal battery capacity in both cases, more energy is required to traverse in rough surface. Nevertheless, similar change of travel time is noticed in both blue and red curves which has parity with the discharge profile of batteries. Thus, the travel time of same arc in different conditions of floor demonstrate that travel time can reflect not only state of charge of batteries \cite{wafPragna} but also environmental conditions. Hence, it is concluded that the travel time of segments inherently represent the dynamically changing factors.

A good model of travel time which incorporates the historical changes of travel time can generate estimates of future travel time. This modelling of travel time based on its change over time eliminates the need to explicitly model different contributing factors to the change of cost. This is achieved using state-space models for travel time. Kalman filter is used to derive the future estimates of travel time which can reflect future costs of traveling an edge. To check whether Kalman Filter works fine to estimate $X$($i$, $j$), experiments were conducted to make robots go from $i$ to $j$ and vice versa so to know about the quality of the estimates. In these experiments, $X$($i$, $j$, $k$+) are estimated using Kalman Filter and the actual measure of $X$($i$, $j$, $k$), where $k$ stands for the number of estimates. 

\begin{figure}[h]
\centering
\includegraphics[scale = 0.36]{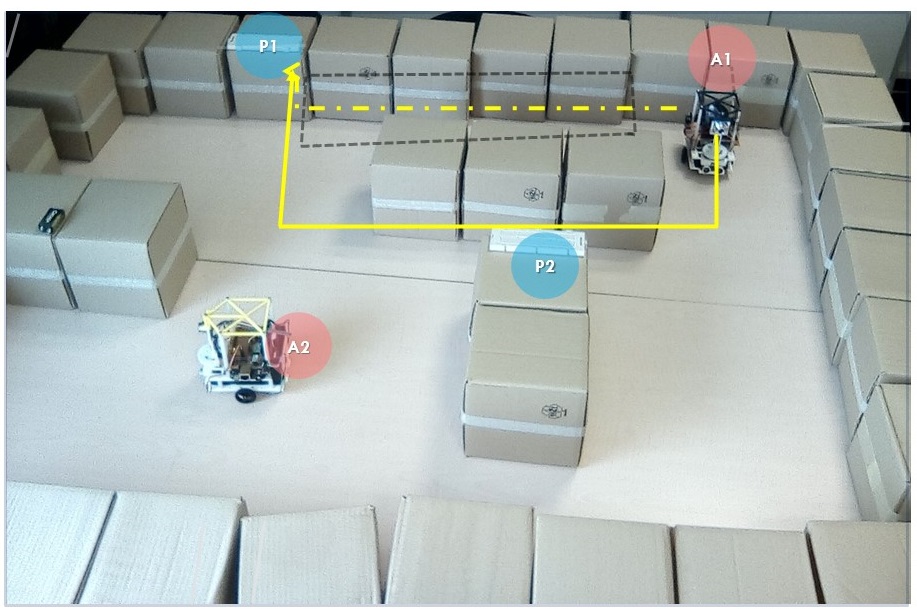}
\caption{An example of more cost efficient path}
\label{fig_exp2}
\end{figure}

The travel time of segments not only depend on change of battery and floor conditions, but also on factors like traffic conditions and behavior of other MRs. So, the travel times represent the real costs as it encompasses internal and external factors. The control at agent level of actuation cannot determine the cost depending on the traffic conditions and other robots' behavior as it does not hold those conditions \cite{etfaPragna}. Hence, even after modelling battery, load and floor conditions, these factors cannot be modelled at agent level. Thus, the true cost cannot be derived at lower level of control. Hence, in this work, cost parameters like travel time is investigated at system level to utilize them efficiently. At this juncture, a more cost efficient path can be computed using the travel times of segments as they are close-to-real costs of traversal. This technique is used as a tool to demonstrate the need and efficacy of travel time in planning, where Dijkstra's planning algorithm is modified to use travel times of edges in order to compute minimum cost paths. In fact, accurate and close-to-real estimated travel times can be used in any path finding algorithm. For example, in Figure~\ref{fig_exp2}, the changed cost of different segments of the floor can be known from the estimation of travel times of the different segments in the floor. When these travel times are used to the path to traverse to $P$ at $t_j$ (when the condition of floor changed), the path marked by solid line is obtained, which is a less cost consuming path to $P$1 than the path marked by dotted lines. 

One of the main advantages of modelling travel time is simplicity in cost computation. Also, only one set of coefficients need to identified rather than three sets ($F$1, $F$2 and $F$). Similar to the above example, a particular (or a set of) costs based on time can be identified for a kind of task, as costs arise due to performance of tasks. The same model of travel time can be used for other kinds of robots. This independence of the model from robot kinematics makes it applicable to a wider range of robots. This is the most significant advantage in respect of applicability. The travel time is calculated considering the difference between the departure from one spot and reaching the next. Thus, the travel time is not dependent on the shape of the segment, rather it depends on the time taken to traverse between any two spots which is joined by the segment. This is another advantage where geometry of the floor do not hinder cost calculation.   

There are two different estimation processes to estimate travel times in this work. At first, the travel times are modeled in a linear state-space model where travel time in current instance depends only on the travel time of previous instance. As a good estimation method to accurately predict travel times requires their histories, observation data for all possible values are obtained offline. As to see how this model and parameter estimation would work in combination with a path planner,  travel times of edges are estimated online during path planning using this model. We compare the cost of optimum paths generated by Dijkstra's algorithm when using regular heuristics as costs ($H$-paths) with the paths obtained from same algorithm using close-to-real travel costs ($R$-paths). The total travelling costs of $H$-paths and $R$-paths are calculated in the same way for comparison. The $X$($i$, $j$, $k$)'s of all consecutive edges are added up, where $i$ and $j$ are two consecutive nodes in the sequence of paths and $k$ is the sequence number. As there is a divergence between $H$-paths and $R$-paths, there is also a divergence in costs which shows that planning using heuristics does not produce optimal paths, not because of the algorithm, but because of costs. From the experiments, it is evident that average total costs of $R$-paths is roughly 5\% less that of $H$-paths. This experiment is described in Section~\ref{exp1}. 

Unfortunately, those total costs of $R$-paths are not computed correctly because $X$($i$, $j$, ($k$+1)) depend on $X$($i$, $j$, $k$) of the same edge whereas edge cost of a path depend on all the costs of the edges previous to current edge as a path is actually a sequence of edge.  In this work, the estimation of travel time is improved considering that current edge cost depend on a set or window of previous edge costs in the sequence and the self-exploration or change of travel time with the progress of operation. For this estimation, a bi-linear state dependent model is used where travel time of current instance depends on a set or window of previous travel times. To account for this, $X$($i$, $j$, $k$) is actually a window of $X$($i$, $j$) values of previous edges in the path, i.e.- $X$($i$, $j$, ($k$-1)), $X$($i$, $j$, ($k$-2))...... $X$($i$, $j$, ($k$-$w$)) where $w$ is the window size. 

In this model, the estimation process start with mean travel time or heuristic travel time and real observations are obtained during the traversal of edge. Thus, observations are collected progressively during MR operation which eliminates the cumbersome process of collecting data offline. The filtering method cannot generate the best estimates at initial few iterations. The estimates get improved over time. In fact, estimating travel time by this method produces paths, named $D$-paths, which are 15\% less cost consuming than the paths obtained by heuristics costs (Section~\ref{exp2}).  

Hence the contribution of this paper is twofold. Firstly, a new method to compute realistic transportation costs in automated logistics is proposed. Travel times are identified as the cost parameter which provide true transportation costs. This work eliminate the need of explicit modelling of the physical and environmental factors as travel time can represent them inherently.
Secondly, the necessity of estimating close-to-real costs and efficacy of the method are shown by using the estimated travel times in route planning for an MR. 
\subsection{Organization of the paper}
The next section elaborates on the related work. Section~\ref{probStatic} and ~\ref{probDynamic} formulates the problem in the light of an internal logistic system with path traversal as a task. Section~\ref{proto} explains the prototype platform and other details for the system used to conduct experiments. The experiments and their results are elaborated in Sections~\ref{procStatic} and ~\ref{procDynamic} and Sections~\ref{resStatic} and ~\ref{resDynamic} respectively. The Algorithm~\ref{staticCost} elaborates on the proposed approach which incorporates modification over Dijkstra's algorithm. Section~\ref{last} concludes with discussions and future directions of investigation.
\section{Related works}
The two prominent planning problems of MRs are autonomous navigation and task scheduling. The autonomous navigation is addressed as a general problem of MR working in any unknown and dynamic environment, while task scheduling is a problem for MRs specifically operating in automated manufacturing units and warehouses. Usually, planning for navigation requires two different but complementary objectives, path planning and trajectory planning \cite{montiel2015path}. There are recent investigations to consider dynamic cost in path planning \cite{cho2017cost}, \cite{fazlollahtabar2018hybrid}, \cite{melo2015multi} yet cost is derived based on the distance between current node to next node or heuristics, not on dynamically changing conditions of environment and battery, though these factors are present on unknown terrain. 
The dynamic factors in automated factories are floor condition, state of battery, mechanical parts of robot while parts like racks, handlers, \textit{et~cetera} remain static mostly. This work addresses to consider these dynamic factors and study their effects in the planning for MR in automated manufacturing and logistics.

In case of proposals dealing with trajectory planning, dynamic cost based on time or energy is either derived out of motor dynamics \cite{kim2014online} or current pose and constant velocity \cite{kollmitz2015time}. In \cite{kollmitz2015time}, cost is not truly represented as it does not consider the change in battery states and environment which induces change in velocity. And in \cite{kim2014online}, the dynamics need to be changed for every new kind of robot model. In current work, travel time is considered as a cost which represents change in states of battery state and floor \cite{wafPragna} and can be derived similarly in any robot. 

Task scheduling, on the other hand, is addressed by introducing several constraints to each task like delivery time, location, transportation capacity of robots, \textit{et~cetera} \cite{veloso2015cobots}. Scheduling addresses to accomplish each task within the specified time taking into account all the constraints. In this context, this work proceeds one step further to estimate these necessary completion times for each task considering the state of charge of batteries and environmental conditions, so that minimum cost in terms of energy and time is expended to accomplish each task. In this work, a new method is proposed to find costs for performing tasks like traversing between spots in order to decide optimally. Here, path planning is considered as an example of planning It is done for a single MR and costs incurred in traversing are predicted by estimating travel times between the spots. Few works on general problem of road-map generation for MRS in logistics have also considered cost as dynamic \cite{digani2015ensemble}, however, it is based on the Euclidean distance.

To best of author's knowledge, travel time is not considered as a cost factor in MRS. 

\section{Prototyped Internal Transportation System}
\label{proto}
\begin{figure}[h]
\centering
\includegraphics[width=0.45\textwidth]{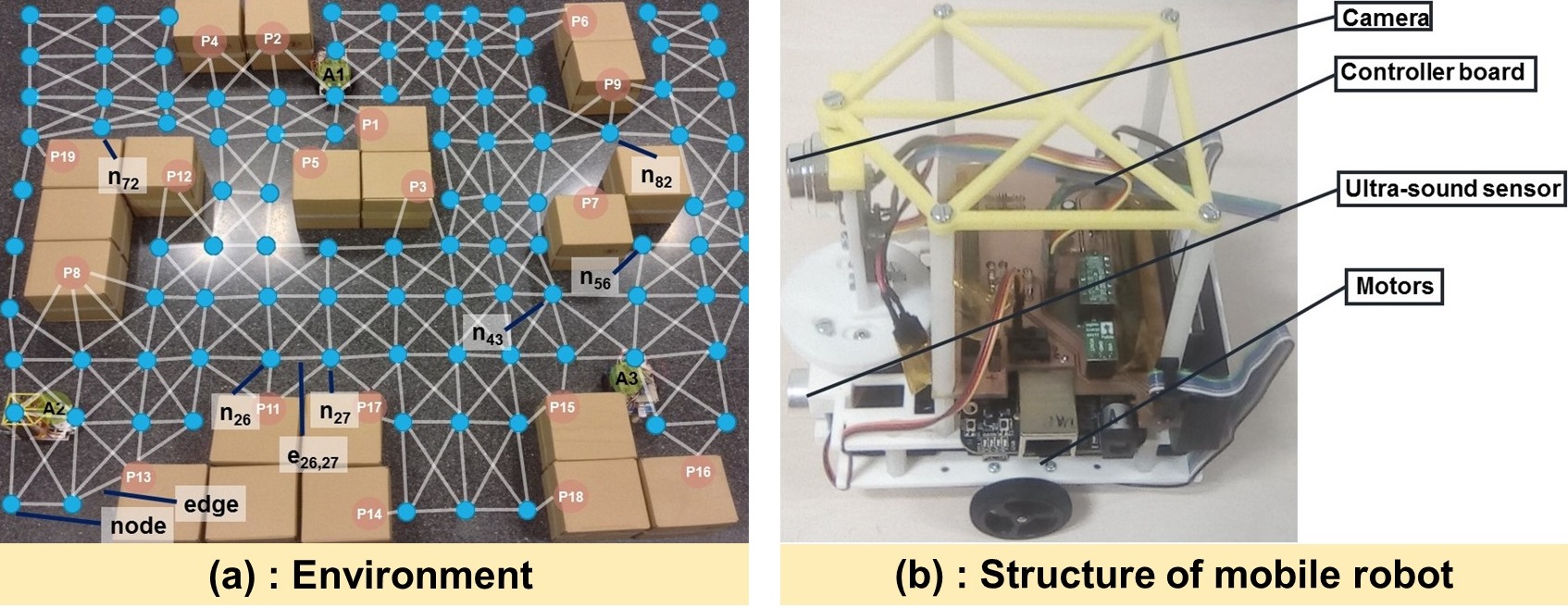}
% where an .eps filename suffix will be assumed under latex, 
% and a .pdf suffix will be assumed for pdflatex; or what has been declared
% via \DeclareGraphicsExtensions.
%\rule{1cm}{1cm}
\caption{Scale downed prototype platform}
\label{envrn&map}
\end{figure}

\begin{figure*}
\includegraphics[scale = 0.378]{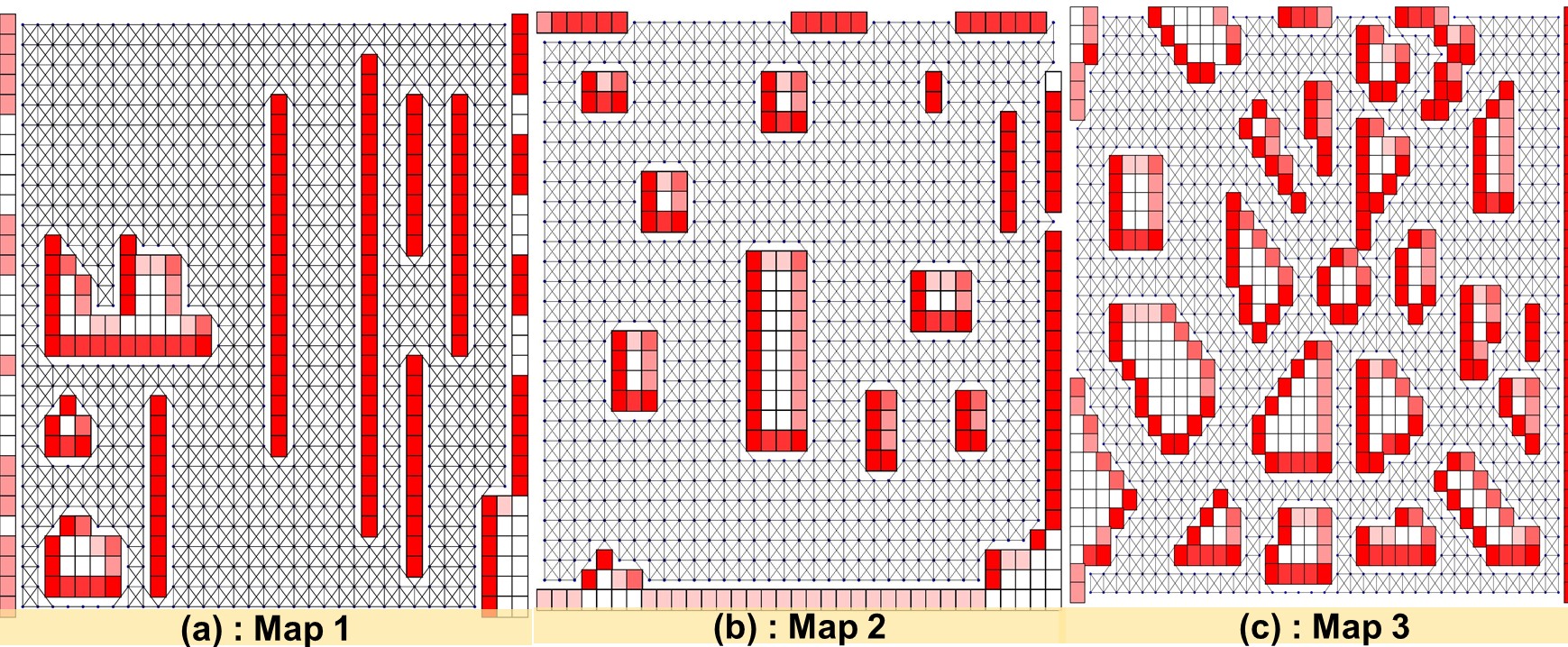}
\caption{Three representative topological maps}
\label{3maps}
\end{figure*}

A prototype scaled down internal transportation system is developed with all essential constituting parts like MRs, tasks, controller architecture and the environment adhering to minute details. The floor is described by means of a topology map $\mathcal{G} = \{\vee,\varepsilon\}$, where each port and bifurcation point corresponds to some node $n_r$ $\in \vee$ and each link between two nodes forms an edge $a_{e,f}\in \varepsilon$. Part (a) of Figure~\ref{envrn&map} depicts a portion of the whole prototype, 
where, notation like $n_{26}$ designates a node and $a_{26,27}$ a edge. Topology maps are generated taking reference from the grid map generated by results of Simultaneous Localisation and Mapping (SLAM) in \cite{beinschob2015graph} based on a simple assumption, that each free cell in the grid map corresponds to a node in the graph. The SLAM from \cite{beinschob2015graph} produces mapping and localization of the Coca-Cola Bilbao plant with good accuracy and thus can be treated as a map of the real plant. 
A selected representative portion from each of the three sections of of Coca-Cola Iberian Partners in Bilbao, Spain are extracted to form three topological maps. They are provided in Figure~\ref{3maps}. Part (a) of Figure~\ref{3maps} illustrates Map~1 which is a representative of winding racks in the warehouse facility, Part (b) shows Map~2 which represents randomly placed racks and Part (c) shows Map~3 which represents racks organized in a hub. 
%\lipsum[1-2]

The scaled robots are built controller, ultrasound sensor and a camera, as illustrated in Part (b) of Figure~\ref{envrn&map}. The DC servo motors drive the wheels of the MR and Li-ion batteries energize them. Each MR has its individual controller in decentralized architecture \cite{ribas2013agent, ribas2009approach}. 

The robots carry out the task of traversing paths to carry materials. This is done by traversing from one spot to another, expressed as nodes. 

\section{Experiment I: Static estimation and using the estimates in route planning}
\label{exp1}
\subsection{Problem definition}
\label{probStatic}
\begin{figure}[h]
\centering
\includegraphics[width=0.45\textwidth]{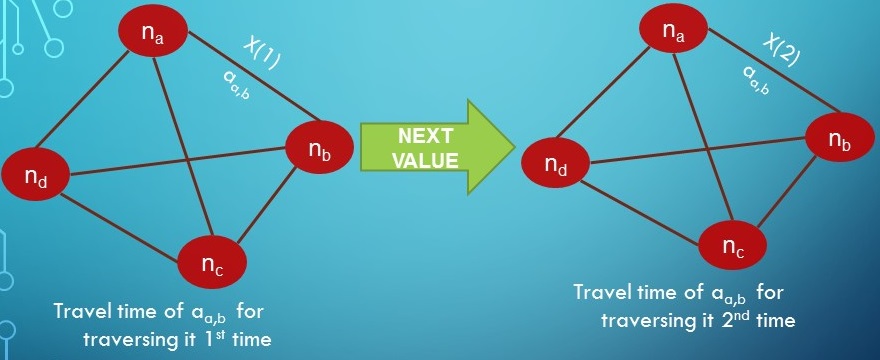}
% where an .eps filename suffix will be assumed under latex, 
% and a .pdf suffix will be assumed for pdflatex; or what has been declared
% via \DeclareGraphicsExtensions.
%\rule{1cm}{1cm}
\caption{Travel time of static estimation}
\label{staticTTPic}
\end{figure}

The problem is to find a suitable model to estimate the travel time of traversing between nodes. In this work, the focus is on one MR. First, $X${$i$,$q$} denote general travel time of a single robot for an edge $a_{i,j}$ between any two nodes $i$ and $j$ (Section~\ref{intro}). For simplicity, $X${$i$,$q$} is also expressed as $X$. The travel time of a particular edge depends on the travel time of that edge at the previous instance. For example, in Figure~\ref{staticTTPic}, travel time for edge $a_{a,b}$ is $X$(1) when it was traversed for first time. Then, when $a_{a,b}$ is traversed the second time, then travel time becomes $X$(2) and so on. Thus, travel time of an edge becomes the function of the number of times that edges is traversed. This is designated as $X$($k$) where $k$ is the number of times that edge is traversed. This is modeled using a state-space model, discussed in Section~\ref{procStatic}. 
\subsection{Procedure}
\label{procStatic}
The first experiment models travel time with the hypothesis that The travel time of a particular edge depends on the travel time of that edge at the previous instance. Based on this, the state-space equations~\ref{state_eqn} and \ref{obs_eqn} models the travel times. 
\begin{align} 
\label{state_eqn}&X(k)= X(k-1) + \omega(k)\\
\label{obs_eqn}&Y(k)= X(k) + \eta(k)
\end{align}
The state vector in equation~\ref{state_eqn} is a single variable $X$ which depends on the number of times the robot has traversed the edge, $k$. Hence, travel cost of an edge is found by estimating its $X$. $Y$($k$) in equation~\ref{obs_eqn} is the observation variable for $X$. This model involves two error terms $\omega$($k$) and $\eta$($k$) which are independent and normally distributed. According to equations~\ref{state_eqn} and \ref{obs_eqn}, $X$($k$) of an edge depends only on the travel time of the edge at the previous instance, i.e.-$X$($k$-1) and observation value of $X$ at $k$. The observation data for all possible $X$s for all $k$s are recorded offline. The travel times for three different length of arcs in all three maps (Figure~\ref{3maps}) and four different conditions of surface till complete exhaustion of batteries are observed. This creates the whole set of observation. This is a cumbersome and time consuming process, yet the travel time was modeled in this way because it is simple and estimation can be performed easily using this model. The goal of the experiment is to verify the online estimation of travel times with the observed data and how it effects planning (Section~\ref{resStatic}). Nevertheless, this problem  is mitigated by a non-linear functional model of $X$ in next experiment (Section~\ref{exp2}). The online estimation of travel time is done by Kalman filtering over the model so that $X$($k$) of an edge depends can be estimated based on $X$($k$-1) and observation value of $X$ at $k$. This estimation process is static as it does not consider the change of $X$ for the total elapse of time since start of system (Section~\ref{drawback}). 
Equations~\ref{KF_state1} and \ref{KF_state2} are obtained after applying Kalman Filtering method \cite{kfRibeiro} on equations~\ref{state_eqn} and \ref{obs_eqn}.
\begin{equation}
\begin{aligned}
\label{KF_state1}&\hat{X}^{-}(k)= \hat{X}(k-1)  \\
&P^{-}(k)= P(k-1) + \sigma^2_\omega 
\end{aligned}
\end{equation}
\begin{align}
\label{KF_state2}
&K(k)=P^{-}(k)\big/[P^{-}(k)+\sigma^2_\eta] \nonumber \\
%P_{j\mid{j-1}}[P_{j\mid{j-1}}+\sigma^2_\eta]^{-1}]\\
&P(k)=P^{-}(k)-[{P^{-}(k)}^2\big/[P^{-}(k)+\sigma^2_\eta]]\\ \nonumber
%P_{j\mid{j-1}} - P_{j\mid{j-1}}^2[P_{j\mid{j-1}}\sigma^2_\eta]^{-1}]\\
&\hat{X}(k)=\hat{X}^{-}(k)+[P^{-}(k)\big/P^{-}(k) 
+\sigma^2_\eta]*\omega(k)\\ \nonumber
&\text{where,}\qquad  \omega(k) = [Y(k)-\hat{X}^{-}(k)] 
\end{align}
%\end{align}\nonumber
$\hat{X^{-}}(k)$ produces the apriori value of $X$ and $P^{-}$ produces the associated covariance, $\sigma^2_\omega$ being the co-variance of process noise $\omega$($k$). $\hat{X}$($k$) provides the predicted estimate of $X$($k$), as $\hat{X^{-}}$($k$) is corrected in equation~\ref{KF_state2} with the help of Kalman Gain $K$($k$). $P^{-}$($k$) provides the associated co-variance matrix, $\sigma^2_\eta$ being the co-variance of the observation noise $\eta$($k$). The initiañl conditions are given by equations~\ref{xinitial} and ~\ref{Pinitial}. 
\begin{align}
\label{xinitial}
&\hat{X}(0)= E[X(0)]\\
%\end{align}
%\begin{align}
\label{Pinitial}
&P(0)= E[(X(0)-E[X(0))(X(0)-E[X(0))^T]
\end{align}
Thus, Kalman filtering produces the estimated travel times based on equations~\ref{KF_state1} and \ref{KF_state2}. 

\begin{figure}[t]
\centering
\includegraphics[scale = 0.35]{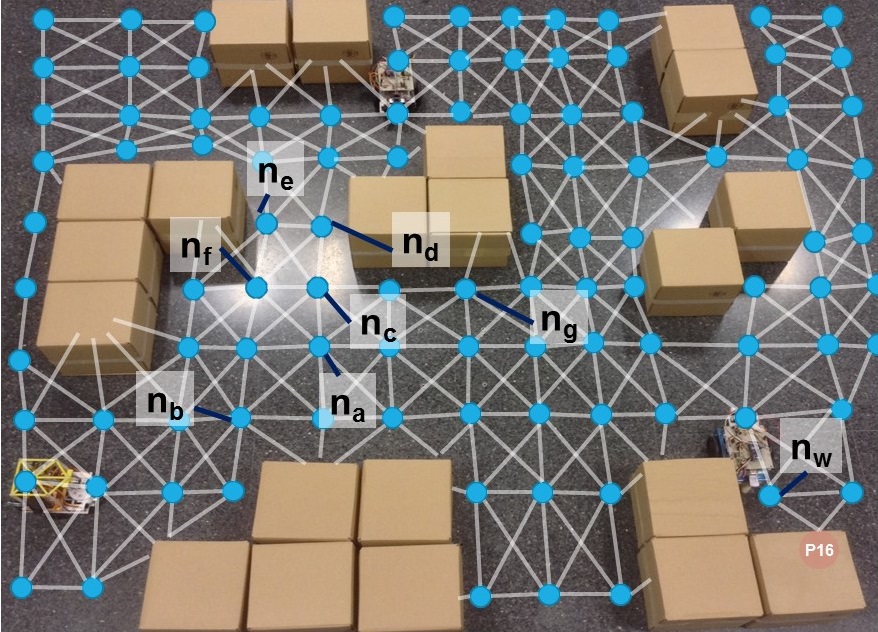}
\caption{Sample run of route computation}
\label{samplerouteplan}
\end{figure}

These estimated travel times are the main instruments to decide a path. Actually, a deterministic path planning algorithm , i.e.-Dijkstra's algorithm is fed with the estimated travel times to determine the least time consuming path, eventually this path is the optimum becomes it incurs least cost in terms of battery and floor utilisation. The reasons for using Dijkstra's algorithm are its simplicity and deterministic nature. The goal of this work is to show that close-to-real cost actually helps in obtaining least cost path. Thus, a simple and deterministic path plannher is deployed to proof the efficacy of estimated travel costs.  While computing path, Dijkstra's algorithm use estimated travel times as the weights of edges to decide the next edge in the path. For example, a sample route computation is illustrated in Figure~\ref{samplerouteplan}. Let $n_a$ be source and $n_w$ destination at $P$16. 
So, path computation using Dijkstra's algorithm starts at $n_a$ with its neighbors $n_b$, $n_c$ and $n_d$. So, $X_{a,b}$, $X_{a,c}$, $X_{a,d}$ are required to be estimated at $k$, when $k$ is 1 as this will be first edge being traversed. We use equation~\ref{KF_state1} to obtain $\hat{X^{-}}$(1) for $X_{a,b}$, $X_{a,c}$, $X_{a,d}$ separately depending on $X$(0) using equation~\ref{xinitial}. Similarly, we get separate $P^{-}$(1) using equation~\ref{Pinitial}. Next, we obtain $\hat{X}$(1) (estimate) and $P$(1) for $X_{a,b}$, $X_{a,c}$, $X_{a,d}$ using equation~\ref{KF_state2}. 
Comparison of estimated values of $X_{a,b}$, $X_{a,c}$, $X_{a,d}$ will provide the least cost of traversing from $n_a$ to any of its neighbor. Let, the least cost edge be $a_{a,c}$. So $n_a$ will become the predecessor of $n_c$, i.e.-to reach $n_c$, the edge should come from $n_a$. When $n_c$ will be explored, the value for $k$ is 2 as $n_c$ has 1 predecessor. The next least cost edge from $n_c$ in the path is required to be known. Thus, $X_{c,e}$, $X_{c,f}$, $X_{c,g}$ needs to be estimated. Thus, observation $Y$($k$) of $X$ at current $k$ is required to estimate $X$. Thus observation values for travel costs of all possible $X$s for all possible $k$s were collected.  Paths are computed consecutively using Dijkstra's algorithm, first using heuristic costs as weights of edges ($H$-paths). Then, the same procedure is repeated with statically estimated travel times as weights of edges ($R$-paths). The $H$-paths and $R$-paths along with their total costs are compared in Section~\ref{resStatic}. 
\subsection{Results} 
\label{resStatic}
Paths are computed repeatedly for 100 times in each topological graph (Figure~\ref{3maps}) from Dijkstra's algorithm for both categories of paths. The choices of source and destination are fed from the decided list of sources and destinations for each call of route computation. 
Section~\ref{intro} elaborated that each edge in the floor is associated with some cost in terms of energy exhaustion and others. The travel time determines the cost of the traversal tasks in the floor (Section~\ref{intro}). A path is a set of connecting edges. A defined path incurs several travel costs for all edges in the path. A path $P$ for a robot is usually expressed in terms of connecting edges as 
\begin{equation}
\label{edgewidP}
\begin{aligned}
P = \langle a_{a,b}, a_{b,c}, a_{c,d}, a_{d,e}, ...........\rangle
\end{aligned}
\end{equation}
Hence, the total travel cost of that path is the total travel time of all edges forming the path. Thus, sum of all travel costs of all edges determines the total travelling cost of the path $P$. This is given by $C_P$ in equation~\ref{costofP}, where, $n_i$ is the source node and $n_d$ is the destination node. 
\begin{equation}
\label{costofP}
\begin{aligned}
C_P = \sum_{n_s}^{n_d} X_{p,q}
\end{aligned}
\end{equation}

\begin{figure}[h]
\centering
\includegraphics[width=0.45\textwidth]{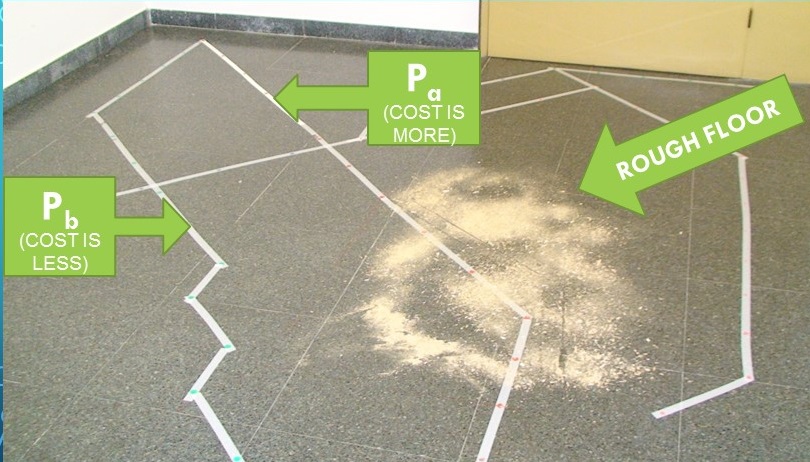}
% where an .eps filename suffix will be assumed under latex, 
% and a .pdf suffix will be assumed for pdflatex; or what has been declared
% via \DeclareGraphicsExtensions.
%\rule{1cm}{1cm}
\caption{Comparison of path costs}
\label{compPathCosts}
\end{figure}

\begin{figure}[h]
\centering
\includegraphics[width=0.45\textwidth]{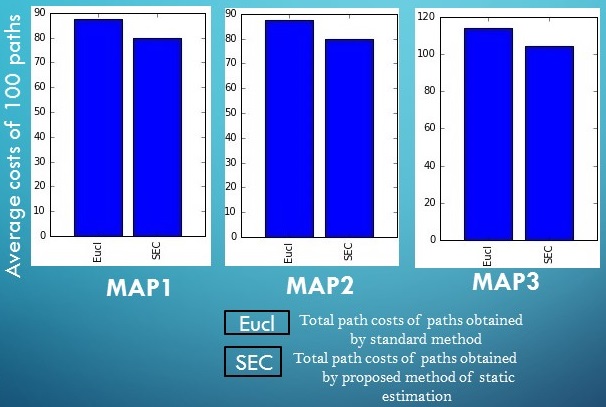}
% where an .eps filename suffix will be assumed under latex, 
% and a .pdf suffix will be assumed for pdflatex; or what has been declared
% via \DeclareGraphicsExtensions.
%\rule{1cm}{1cm}
\caption{Results of static estimation}
\label{tstaticRes}
\end{figure}

In experiment~\ref{procStatic}, total costs of $H$-paths and $R$-paths are computed according to equation~\ref{costofP}. The $C_P$ of $H$-paths are computed by replacing the edge weights by the real edge costs obtained from estimation of travel times.  Then, the total path costs of $H$-paths are compared with $R$-paths. For example, in Figure~\ref{compPathCosts}, $P_a$ is an example of $H$-paths  and $P_b$ is an example of $R$-paths. As heuristic costs do not encompass the states of floor and battery, the edges of $P_a$ passes through rough zone of the floor. The real total cost of traversing $P_a$ is obtained by calculating the sum of the travel times of it's constituting edges (Equation~\ref{costofP}), obtained from estimation.
The average of total travel costs of all 100 of $H$ and $R$-paths are calculated separately. %The average of total travel cost of $H$-paths never change with increase in number of total paths,i.e.-average total costs of sets of 20, 40, 60 and 80 $H$-paths remain same as heuristic weights do not change over time and does not reflect the true cost of traversal. Also, average of total travel cost of set of 20, 40, 60 and 80 $R$-paths do not change significantly with increase of total number of iterations of path computation. This happens because there is lack of variation in data. 
The Figure~\ref{tstaticRes} plots the average of total travel costs of 100 paths from both $H$ and $R$-paths in three maps (Figure~\ref{3maps}). The vertical bars of \textit{Eucl} and \textit{SEC} show the average of total travel costs of 100 $H$ and $R$-paths in all three maps, respectively. 

Vertical bar \textit{SEC} shows that average of total path costs of 100 $R$-paths is 5\% less in case of Map~2 and Map~3 and 2\% less in Map~1 than that of 100 $H$-paths. 

The static estimation process is executed to corroborate that travel time can be modelled, without the process of deriving it through battery, load and environment modelling and that its online estimation is possible. This experiment also verifies that weights of edges can be estimated as travel time online during exploration of Dijkstra's algorithm using a state-space model. Moreover, these estimates can generate path with less total costs than paths obtained through heuristics cost. Also, it is verified that the estimated values of $X$ are correct and real through this experiment, as the values can be compared to real observations. The paths obtained had less total cost than that of paths obtained through heuristics cost. This assures that travel time represents close-to-real costs and hence it effects planning. Despite involving the cumbersome process of gathering observation data, this static experiment has these above positive results about online estimation of travel time and its effect on planning. 
%Despite having the drawbacks,  Moreover, the cost of traversing each edge is obtained by estimating its travel time based on this model. These static estimates are used as weights of edges to find paths. 
\subsection{Drawback of static estimation}
\label{drawback}
\begin{figure}[h]
\centering
\includegraphics[width=0.45\textwidth]{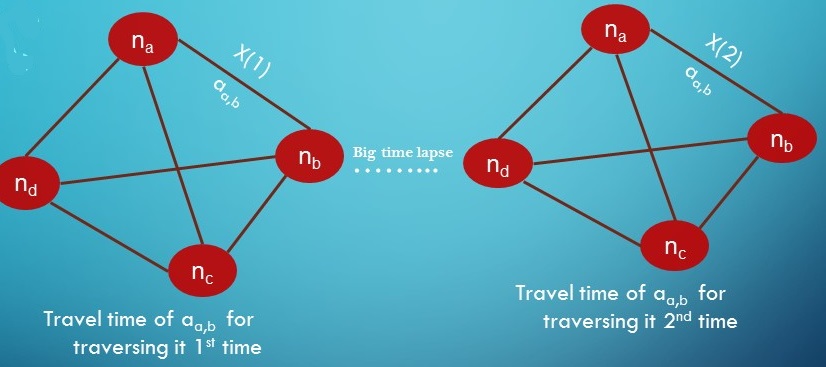}
% where an .eps filename suffix will be assumed under latex, 
% and a .pdf suffix will be assumed for pdflatex; or what has been declared
% via \DeclareGraphicsExtensions.
%\rule{1cm}{1cm}
\caption{Drawback of static estimation}
\label{staticDrwaback}
\end{figure}

In the model of travel time (equations~\ref{state_eqn} and~\ref{obs_eqn}), the estimated value of $X$($k$+1) depends only on $X$($k$) and the observation of $X$ at ($k$+1). This is the drawback in the model as, in reality, it depends on $X$s for all the previous edges in the path and its own variation over the time.
%Still,  as $X$ for any edge depends on all the $X$s of the previous edges which are chosen to form the path. 
The reason being the discharge of batteries and (or) possible change of environment. In this process $X$ is estimated without considering its variation with the total elapse of time from start of system and thus it is static estimation. For example, in Figure~\ref{staticDrwaback},
when edge $a_{a,b}$ was traversed for
first time, the travel time for $a_{a,b}$ was $X$(1). In reality, the robot traverses other edges before traversing $a_{a,b}$ again. there are other nodes and time lapse between two traversals of $a_{a,b}$. Then, when $a_{a,b}$ is traversed the second time, then
travel time becomes $X$($n$) where $n$ $>$ 1  and depends on number of edges being traversed before travelling $a_{a,b}$. 
Also, the estimated value of $X$ at current instance ( i.e. at ($k$+1)) depends only on previous $X$ ( i.e. at ($k$)) and the observation of $X$ at ($k$+1). Thus, observation for all possible $X$s at possible instances are needed to estimate in this process. This is not only cumbersome but also impractical to gather such huge amount of observations. These two drawbacks are solved in the dynamic estimation, whose problem formulation is described in Section~\ref{probDynamic}.  %Paths are obtained continuously for 100 iterations using Dijkstra's algorithm using first heuristics cost based on distance and then this estimated travel time values as weights for edges. The average total costs of two categories of paths are compared. %These experiments and results are explained in Section~\ref{exp1}. 
\section{Experiment II: Dynamic estimation and using the estimates in route planning}
\label{exp2}
The drawbacks of static estimation is elaborated in Section~\ref{drawback} and the need of dynamic estimation is motivated in Section~\ref{probDynamic}. The next section formulates the problem arising out of drawbacks of static estimation and describes the solution to alleviate them. 
\subsection{Dynamic estimation}
\label{probDynamic}
\begin{figure}[h]
\centering
\includegraphics[width=0.45\textwidth]{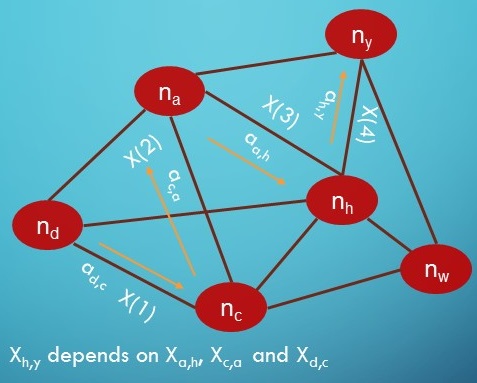}
% where an .eps filename suffix will be assumed under latex, 
% and a .pdf suffix will be assumed for pdflatex; or what has been declared
% via \DeclareGraphicsExtensions.
%\rule{1cm}{1cm}
\caption{Travel time of paths}
\label{trueTT}
\end{figure}

A defined path $P$ for a robot is usually expressed in terms of connecting edges (Sections~\ref{resStatic}) as 
\begin{equation}
\label{pathVector}
\begin{aligned}
P = \langle a_{a,b}, a_{b,c}, a_{c,d}, a_{d,e}, ...........\rangle
\end{aligned}
\end{equation}
The robot first traverses $a_ {a,b}$ and then subsequently all other edges after $a_ {a,b}$. The first edge may (or may not ) be traversed after traversing all the edges in the path. During this time, battery gets exhausted and floor condition may (or not) change. When $a_{a,b}$ is being traversed for second time, it's $X$ depends on all the $X$s of the previous edges in the path. For example, in Figure~\ref{trueTT}, for traversing from node $n_d$ to node $n_y$, there are many intermediate nodes $n_c$, $n_a$ and $n_h$. 
Hence, $X_{h,y}$ depends on $X_{a,h}$, $X_{c,a}$ and $X_{d,c}$. As, robot traverses $a_{d,c}$ first, then $X_{d,c}$ is the $X$(1). Similarly, $X_{c,a}$ is $X$(2), $X_{a,h}$ is $X$(3) and $X_{h,y}$ is $X$(4). Thus, $X$ becomes the function of number of times robot crosses edges to reach destination. This number is designated as $m$. A window of current $X$ and a fixed number (let $j$) of values of previous $X$s are used to form a state vector. The previous values of $X$s are the travel times of those edges which are already considered to form the path. Also, state vector contains an exploration variable $\xi$($k$). A fixed window of values of $\xi$($k$) of same size $j$ is considered in the state vector. Thus, the state vector contains both parts, the change of $X$ over time and the $X$s of previous edges. The state vector (let $s$) is estimated on every $m$ and $s$($m$+1) is formed. $X$($m$+1) is one element of $s$($m$+1). Thus, $X$($m$+1) is formed for every $m$. Here, the current value of $X$ is estimated depending on the previous $X$s i.e.-travel times of edges along with the variation of exploration of $X$ due to elapse of time. Thus, $X$ values are dynamically estimated considering its variation over elapse of time. Moreover, the model is allowed to gather the possible values of $X$ itself from the beginning of first call of path planning and use these values to estimate current value. Observations of all possible $X$($m$) for all possible $m$ are not needed in the latter. This experiment is elaborated in next section (Section~\ref{procDynamic}). 
\subsection{Procedure}
\label{procDynamic}
 In this section, the procedures and results of dynamic estimation of travel times are described. 
The bi-linear model \cite{priestley1988}, provided in equation~\ref{bilinear} is used to model the change of travel costs depending upon all the previous travel costs. $X$ is formed as a function of its past histories over $k$, considering the progressive change $\xi$ with respect to $k$.   
\begin{align}
\label{bilinear}
X(m)+a_1X(m-1)+.....+a_jX(m-j)\\ \nonumber
= \xi_m +b_1\xi(m-1)+...+b_l\xi(m-l) \\ \nonumber
+\sum\sum c_{rz}\xi(m-r)X(m-z) + \omega(m-1) \nonumber
\end{align}
The double summation factor over $X$ and $\xi$ in the above equation ~\ref{bilinear}
provides the nonlinear variation of $X$ due to state of batteries and changes in environment. The state space form of the bi-linear model is given in equations~\ref{bStateEqn} and \ref{bObsEqn}.
The equation~\ref{bStateEqn} is the state equation and the state vector $s$($m$) is of the form $(1,\xi(m-l+1),....,\xi(m),X(m-j+1),......,X(m))^T$. 
Here, $j$ and $l$ denote number of previously estimated $X$s and previous innovations of $X$ respectively. The term $regression\_no$ denotes the values of $j$ and $l$ and is chosen as a design parameter. The $regression\_no$ is increased from 2 to 9 and the effects on total edge travel cost of paths is demonstrated in Section~\ref{resDynamic}. %, its effect on the  
\begin{align}
\label{bStateEqn}
s(m) = F(s(m-1))s(m-1) + V\xi(m)+G\omega(m-1)
\end{align}
\begin{align}
\label{bObsEqn}
Y(m) = Hs(m-1) + \xi(m)+ \eta(m)
\end{align}
%\end{aligned}
%\end{equation}
%\begin{equation}
%\begin{aligned}
%\end{aligned}
%\end{equation}
The state transition matrix $F$ in the equation~\ref{bStateEqn} has the form of
\[
F =
\adjustbox{width=0.7\linewidth}{
$\begin{bmatrix}
    1\mspace{18.0mu}0\mspace{18.0mu} 0\mspace{18.0mu}\dots\mspace{18.0mu}0\mspace{18.0mu}\vdots\mspace{18.0mu}0\mspace{18.0mu} 0\mspace{18.0mu}\dots\mspace{18.0mu} 0\mspace{18.0mu}0\\
    0\mspace{18.0mu}0\mspace{18.0mu} 1\mspace{18.0mu}\dots\mspace{18.0mu}0\mspace{18.0mu}\vdots\mspace{18.0mu}0\mspace{18.0mu} 0\mspace{18.0mu}\dots\mspace{18.0mu} 0\mspace{18.0mu}0\\
    0\mspace{18.0mu}0\mspace{18.0mu} 0\mspace{18.0mu}\dots\mspace{18.0mu}1\mspace{18.0mu}\vdots\mspace{18.0mu}0\mspace{18.0mu} 0\mspace{18.0mu}\dots\mspace{18.0mu} 0\mspace{18.0mu}0\\
    0\mspace{18.0mu}0\mspace{18.0mu} 0\mspace{18.0mu}\dots\mspace{18.0mu}0\mspace{18.0mu}\vdots\mspace{18.0mu}0\mspace{18.0mu} 0\mspace{18.0mu}\dots\mspace{18.0mu} 0\mspace{18.0mu}0\\
    \vdots\mspace{18.0mu}\vdots\mspace{18.0mu}\vdots\mspace{18.0mu}\dots\mspace{18.0mu} \vdots\mspace{18.0mu}\vdots\mspace{18.0mu}\vdots\mspace{18.0mu} \vdots \mspace{18.0mu}\dots\mspace{18.0mu} \vdots\mspace{18.0mu} \vdots\\
    0\mspace{18.0mu}0\mspace{18.0mu} 0\mspace{18.0mu}\dots\mspace{18.0mu}0\mspace{18.0mu}\vdots\mspace{18.0mu}0\mspace{18.0mu} 1\mspace{18.0mu}\dots\mspace{18.0mu} 0\mspace{18.0mu}0\\
    0\mspace{18.0mu}0\mspace{18.0mu} 0\mspace{18.0mu}\dots\mspace{18.0mu}0\mspace{18.0mu}\vdots\mspace{18.0mu}0\mspace{18.0mu}0 \mspace{18.0mu}1\mspace{18.0mu}\dots\mspace{18.0mu}0\\
    0\mspace{18.0mu}0\mspace{18.0mu} 0\mspace{18.0mu}\dots\mspace{18.0mu}0\mspace{18.0mu}\vdots\mspace{18.0mu}0\mspace{18.0mu} 0\mspace{18.0mu}0\mspace{18.0mu}\dots\mspace{18.0mu} 1\\
    \mu\mspace{18.0mu}\psi_l\mspace{18.0mu}\psi_{l-1}\mspace{18.0mu}\dots\psi_1\vdots-\phi_j-\phi_{j-1}\dots-\phi_1
\end{bmatrix}$}
\]
The number of rows of $F$ is given by (2*$regression\_no$ + 1). 
%The matrix $F$ contains coefficient terms as $\psi$ , $\phi$ , $\mu$, \textit{et~cetera}.

The $\psi$ terms in $F$ are denoted as in equation~\ref{psiterms}
\begin{equation}
\label{psiterms}
\begin{aligned}
\psi_l =b_l+ \sum_{i=1}^{l} c_{li}X(m-i)\\
\end{aligned}
\end{equation}
All the $\phi$ terms in $F$ are constants. The term $\mu$ is the average value of $X$ till $k$. %Thus, the state transition matrix $F$ depends on the travel times of fixed number of previously traversed edges. 
Also, the matrix $V$ in \ref{bStateEqn} is denoted as 
\[
V^T = 
\begin{bmatrix}
    0 & 0 & 0 & \dots & 1 & \vdots & 0 & 0 & \dots & 1
\end{bmatrix}
\]
The number of rows of $V$ is again given by (2*$regression\_no$ + 1). The matrix $H$ in \ref{bObsEqn} is denoted as 
\[
H = 
\begin{bmatrix}
    0 & 0 & 0 & \dots & 0 & \vdots & 0 & 0 & \dots & 1
\end{bmatrix}
\]
\[
G^T = 
\begin{bmatrix}
    0 & 0 & 0 & \dots & 0 & \vdots & 0 & 0 & \dots & 1
\end{bmatrix}
\]

The equation~\ref{bObsEqn} is the observation equation. This model allows to gather information about $X$ for different arcs in the map gradually with time during operation. After start of computing a path, the real travel time of edges are recorded when the MR actually traverses it. This travel times of edges are used as the observation values for the next call of path planning. Thus observation values of travel times of each edge is grown during run-time. 
%This travel times of edges are used as the observation values for the next call of path planning. Thus observation values of travel times of each edge is grown during run-time.

Kalman filtering is applied on the state-space model (equations~\ref{bStateEqn} and \ref{bObsEqn}) resulting in equations~\ref{KF1dynamic} and \ref{KF2dynamic} to estimate $s$ repeatedly to obtain $X$ for the connecting edges at each node to compute path using Dijkstra's algorithm.  
\begin{align}
\label{KF1dynamic}
%\begin{aligned}
%\begin{split}
&\hat{s^{-}}(m) = F(s(m-1))s(m-1) + V\xi(m)+G\omega(m-1)\\ 
&\hat{P^{-}}(m) = F(s(m))P(m-1)F^T(s(m-1))+ Q(m-1) 
%\end{split}
%\end{equation}
%\begin{equation}
\end{align}
In equation~\ref{KF1dynamic}, $\hat{s^{-}}$($m$) provides the apriori estimate of $s$. $\hat{P^{-}}$ provides the associated covariance matrix where $Q$($m$-1) provides the covariance for the process noise $\omega$($m$-1).
\begin{align}
\label{KF2dynamic}
%\begin{split}
&K(m)= \hat{P^{-}}(m)H^T[H\hat{P^{-}(m)}H^T+R(m)]\\ \nonumber
&\hat{s}(m) = \hat{s^{-}}(m)+K(m)[Y(m)-H\hat{s^{-}}(m)]\\ \nonumber
&P(m) = [I - (K(m))H]\hat{P^{-}}(m) 
\end{align}
In equation~\ref{KF2dynamic}, $K$($m$) is the Kalman gain, $R$($m$) being the covariance of observation noise $\eta$($m$). $\hat{s}$($m$) provides the estimated state vector $s$ at $m$. 
%From this we can obtain estimate of $X$ for edges. 
\begin{align}
\label{sinitial}
&\hat{s}(0)= E[s(0)]\\ 
\label{P2initial}
%\begin{aligned}
&P(0)= E[(s(0)-E[s(0))(s(0)-E[s(0))^T]
\end{align}

\begin{figure}[t]
\centering
\includegraphics[scale = 0.35]{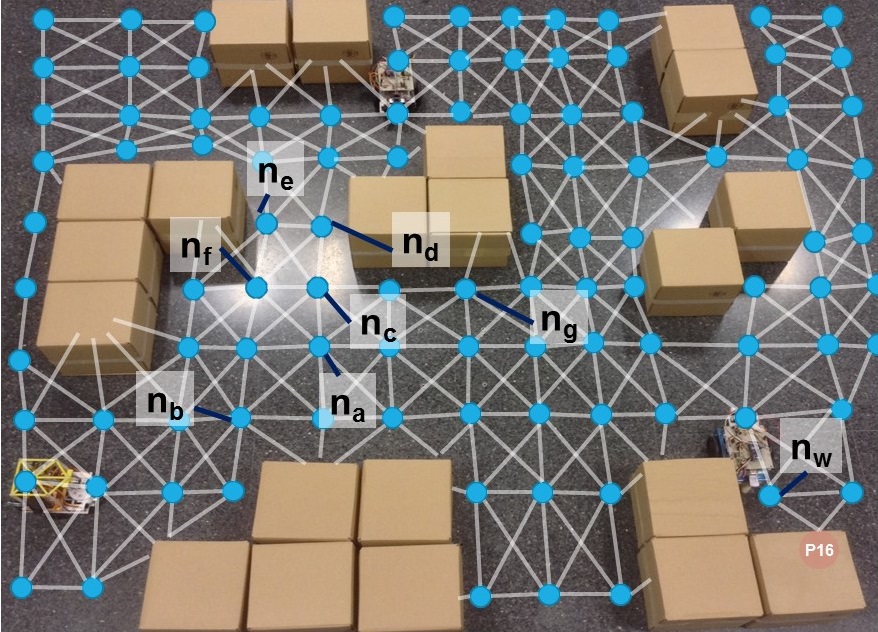}
\caption{Sample run of route computation}
\label{samplerouteplan1}
\end{figure}

Figure~\ref{samplerouteplan1} is used again, as in Section\ref{procStatic}, to explain the dynamic estimation process.
The path computation starts at $n_a$. Let the values of $j$ and $l$ are equal which is 2. At start, $k$ is 1. Now $s$ cannot be formed as minimum 2 previous travel costs are needed. Exploration proceeds with average travel cost for the edges. When $n_c$ needs to be explored, value of $k$ becomes 2 as one travel cost has been known connecting $n_c$ to its predecessor $n_a$. $s$ can be now formed as $X$(1) is known. Again, $n_a$ is the source and so $X$(0) is 0. $\xi$ is assumed to be $N$(0.1,0.1). At $k$ =2, $s$(1) takes the form $(1, \xi(0),\xi(1), X(0), X(1))^T$. Equation~\ref{KF1dynamic} and \ref{KF2dynamic} are used to estimate $s$(2) separately for all edges arising out of $n_c$ to obtain $X$ for each edge. From equations~\ref{sinitial} and \ref{P2initial}, $s$(0) and $P$(0) can be obtained. Let at $n_g$, $k$ = 4. Hence, $X$(3) will be travel cost from $n_e$ (predecessor of $n_g$) to $n_g$, $X$(2) will be travel cost from $n_c$ (predecessor of $n_e$) to $n_e$, $X$(1) will be travel cost from $n_a$ (predecessor of $n_c$) to $n_c$. Thus, $s$(3) = $(1,\xi(2),\xi(3), X(2), X(3))^T$ and $s$(4) = $(1,\xi(3),\xi(4), X(3), X(4))^T$ needs to be computed. This approach is different from Algorithm~\ref{staticCost} in the way the $X$ is estimated. 
\subsection{Results}
\label{resDynamic}
The process of path computation is exactly similar to previous experiment. The paths are obtained in two categories, first with heuristics cost as weights ($H$-paths) and second with dynamically estimated travel times as weights ($D$-paths). In this experiment, total costs of $H$-paths and $D$-paths are computed according to equation~\ref{costofP}. The $C_P$ of $H$-paths are computed by replacing the edge weights by the real edge costs obtained from estimation of travel times.  Then, the total path costs of $H$-paths are compared with $D$-paths. Hence, this experiment differs than the previous experiment in the way travel times are estimated. Here, dynamic estimation of $X$s for relevant edges are done using the bil-inear state space model. The $b$ and $c$ of the model parameters are chosen as normal distribution. Along with the repetitions of path computations, the value $\phi$, mean and covariance of $b$ and $c$ are increased from -0.4 to 0.4 and from -0.2 to 0.2 respectively. Negative values of $\phi$ produced too high estimates while values greater than 0.2 produced negative estimates. Similarly, mean and co-variance values less than 0.1 produce high estimates and more than 0.1 produce negative estimates. Thus, 0.2 is found as the suitable value of $\phi$ and $N$(0.1,0.1) suits for both $b$ and $c$. Moreover, the \textbf{$regression\_no$} are increased from 2 to 9 for each of 20, 40, 60 and 80 repetitive computations. 

\begin{figure}[h]
\centering
\includegraphics[width=0.45\textwidth]{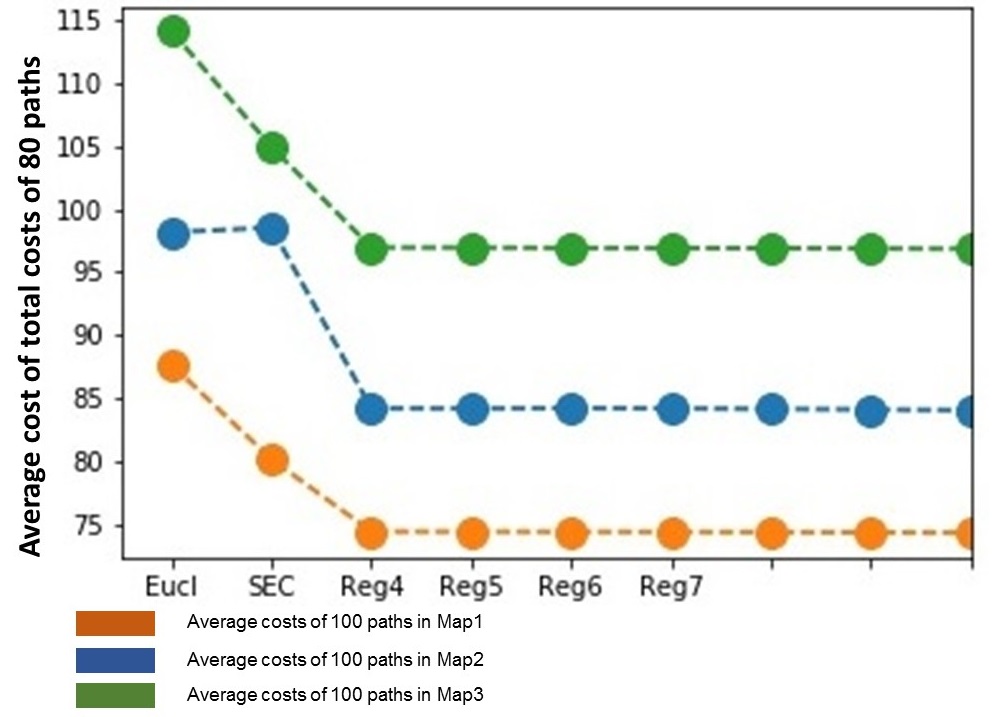}
% where an .eps filename suffix will be assumed under latex, 
% and a .pdf suffix will be assumed for pdflatex; or what has been declared
% via \DeclareGraphicsExtensions.
%\rule{1cm}{1cm}
\caption{Results of dynamic estimation}
\label{tdynamicRes}
\end{figure}

Similar as previous, the total path costs of 20, 40, 60 and 80 bundles of paths are obtained in two categories of paths. 
Also, the average of total travel cost of these bundles in two categories of path do not vary significantly due to increase of iterations. Figure~\ref{tdynamicRes} plots the average of total travel costs of 80 paths from both categories in three maps (Figure~\ref{3maps}). 
The vertical bars of \textit{Eucl} show the average of total travel costs of 20 paths of first category of paths in all three maps. The vertical bars marked from $Reg_2$ to $Reg_9$ represent the average of total path costs for 20 paths in second category. These bars from $Reg_2$ to $Reg_9$ show that they are 15\% less on average than heuristic euclidean cost for all three maps. This difference is increased with the increase of \textit{$regression\_no$}, though the rate of increase is low, as the change of $X$ itself is not broadly spread with standard deviation of 0.219 on average. The average total path cost increases with increase in number of repetitions as edge travel cost increases with elapse of time. 
%The observation $Y$($k$) developed during run-time is considered as signal and the values of $\omega$ are modified to increased the Signal-to-Noise Ratio (SNR) from 10dB to 50 dB along with the repetitions of path planning. The vertical bars marked 10dB, 25dB and 50dB in Figure~\ref{snrPic} plots the average path costs obtained by changing the SNR for each \textit{$regression\_no$}. which shows that with the increase of SNR, the average travel cost decreases.
\subsubsection{Path comparison}
\begin{figure}[h]
\centering
\includegraphics[width=0.25\textwidth]{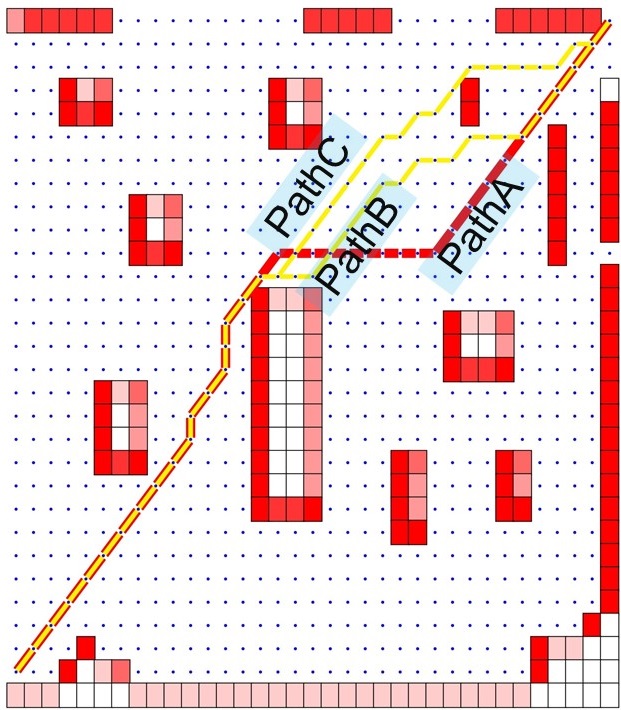}
% where an .eps filename suffix will be assumed under latex, 
% and a .pdf suffix will be assumed for pdflatex; or what has been declared
% via \DeclareGraphicsExtensions.
%\rule{1cm}{1cm}
\caption{Three different paths by three differently obtained travel times}
\label{map2paths}
\end{figure}

Figure~\ref{map2paths} plots 3 single paths Path$A$, Path$B$ and Path$C$ obtained from Dijsktra's algorithm based on heuristic costs, statically estimated and dynamically estimated edge travel costs respectively for the same pair of source and destination nodes in Map~2 including only the variation induced by discharge of batteries. Thus, Path$A$, Path$B$ and Path$C$ are a single example in each $H$-paths, $R$-paths and $D$-paths categories, respectively. 
Here, $P_{A}$, $P_{B}$ and $P_{C}$ are the path vectors (Section~\ref{probDynamic}) for Path$A$, Path$B$ and Path$C$ respectively. Despite having common elements, the three paths are different at the beginning part. Thus, the total travel cost in these 3 paths are different. According to Figure~\ref{tdynamicRes}, average cost of $H$-paths are 15\% more than that of $D$-paths and 5\% more than that of $R$-paths. After obtaining the total travel costs of Path$A$, Path$B$ and Path$C$, it can be stated that, 
\begin{align}
    \sum C_{PB} < \sum C_{PA} by 5\% \nonumber and 
    \sum C_{PC} < \sum C_{PA} by 15\% \nonumber
\end{align}
This also establishes the proposal that heuristics based path planning can underestimate real edge travelling costs and lead to expensive paths. 

\begin{figure}[h]
\centering
\includegraphics[width=0.45\textwidth]{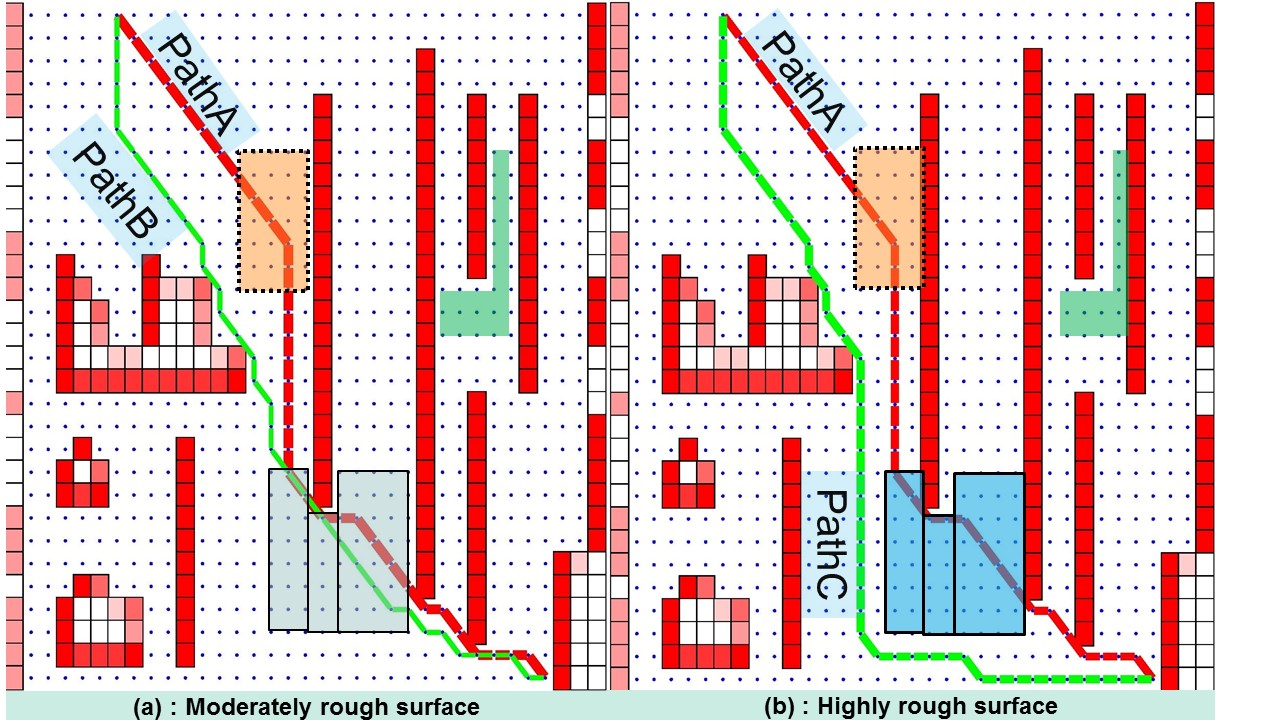}
% where an .eps filename suffix will be assumed under latex, 
% and a .pdf suffix will be assumed for pdflatex; or what has been declared
% via \DeclareGraphicsExtensions.
%\rule{1cm}{1cm}
\caption{Paths differ when conditions change as travel times change too}
\label{map1paths}
\end{figure}

Additionally, Path$A$ and Path$B$ in (a) of Figure~\ref{map1paths} are obtained in Map~1 by heuristic based edge weights and dynamically estimated edge travel costs respectively when floor condition in dotted line zone is moderately rough and solid line zone is lightly rough.  

Path$A$ in (a) contains edges in both rough zones in the floor, while Path$B$ in (a) clearly avoids the zone with moderate roughness, though having few edges in the lightly rough zone. This happens because Dijkstra's algorithm finds that cost incurred in traversing the lightly rough zone to be less than that of the additional edges required to avoid the zone. This proves that modification of Dijkstra's algorithm using dynamically estimated travel cost does not disrupt the computational robustness of the algorithm. Also, when the lightly rough zone is made heavily rough, Path$B$ deviates and Path$C$ is obtained as shown in (b) in Figure~\ref{map1paths}. Thus again, estimated travel times of edges help Dijkstra's algorithm to find a cost effective path. 
\subsubsection{Real cost saving for paths}
In (a) of Figure~\ref{map1paths}, there are total 12 edges from the 2 rough zones comprised in Path$A$. The path cost of Path$A$ obtained using heuristic weights is not the correct one as travel costs of each of these 12 edges are more than assumed. Let, a variable $\delta$ accounts for the additional edge costs in each edge. Path cost of Path$A$ is obtained as 98.210 from results, but in reality path cost of Path$A$ should be (98.210 +12*$\delta$). The value of $\delta$ can never be zero as changes in environment ans batteries will always be present. When more zones will have changed floor conditions, more edges will have increased edge cost. So, the coefficient of $\delta$ will increase and also the true value of travel cost of paths. Thus, the difference between travel costs of paths obtained by heuristic cost and estimated travel time will always increase with the increase of hostility in the environment.

The methodology of using the online estimated travel time in Dijsktra's algorithm is explained in  Algorithm~\ref{staticCost}. In this methodology, first statically estimated travel times (Section~\ref{exp1}) are used and then dynamically estimated travel times (Section~\ref{exp2}) are used. 
\begin{algorithm}
 \caption{Using estimation of travel time in Dijkstra's algorithm}
 \label{staticCost}
 %\begin{multicols}{2}
 \begin{algorithmic}[1]
   %\SetKwInOut{Input}{Input}
    %\SetKwInOut{Output}{Output}
    \Function{initialise\_single\_source}{$\vee,s$}
    \Comment{Where $\vee$ - list of nodes, s - source, returns $d[v]$ - atribute for each node, $\pi[v]$ - predecessor for each node}
    %\underline{initialise\_single\_source} $(\vee,s)$\\%\;
   % \Input{$\vee$-list of nodes, $s$-source node}
   % \Output{$d$[$v$]-attribute for each each node, $\pi$[$v$]-predecessor of each node}
%\Function{Create\_Topo\_Map}{$A,r,c$}\Comment{A-grid map, r-number of rows, c-number of columns}\\
%\pagebreak
%\SetAlgoLined
%\KwResult{Write here the result }
 %totalrowcount := row number of grid map\;
 %totalcolcount := column number of grid map\;
 \For{each $x_i \in V$}
    \State $\pi[x_i] = infinity$
    \State $d[x_i] = NIL$
  \EndFor
    
 %$d$[$s$] = 0\\ 
 %$\pi$[$s$] = NIL
 \EndFunction
  \Function{find\_prev}{$(u,s)$}
    \Comment{input: $u$-current node,$s$-source node, returns: $prev\vee$-predecessor of $u$, $noPred$ -number of predecessors till $s$}
       \State $prev\vee$ = compute predecessor of $u$
       \State $noPred$ = count of predecessors till $s$
        \EndFunction
    \Function{KF}{$(pW,k,Y(k))$}\Comment{input: $pW$-value of travel time at $k$ -1, $k$-instance for estimation, $Y$- observation variable, Returns: $\hat{X}$($k$)-travel cost from $u$ to $v$}        
       % \underline{KF} $(pW,k,Y(k))$\\%\;
    %\Input{}
    %\Output{$\hat{X}$($k$)-travel cost from $u$ to $v$}
    \State Apply KF on state-space model to obtain $\hat{X}$($k$)
    \EndFunction
    \Function{find\_cost}{$u,v,k,pW,Y(k)$}%\;
    \Comment{Input: $u$-current node, $v$- neighbor node, $k$- instance of estimation,$pW$ - cost between prev$u$ and $u$, $Y$($k$) - observation of travel time between $u$ and $v$
    ,Returns:$w$- estimated travel\_time (cost) from $u$ to $v$}
        \State $w$ = $KF$($pW$,$k$,$Y$($k$))
     \EndFunction
     \Function{Relax}{$u,v,w$}%\;
    \Comment{Inputs: $u$-current node, $v$- neighbor node, $w$- estimated travel\_time (cost) from $u$ to $v$, Returns: $d$[$v$]-attribute for each each node, $\pi$[$v$]-predecessor of each node}
    \If{$d$[$v$] $> d$[$u$] + $w$($u,v$)}
      \State $d[v] = d[u] + w(u,v)$
      \State $\pi[v] = u$
      \EndIf    
    \EndFunction
\Function{Main}{$\vee, \varepsilon, Y, s$}%\;
    \Comment{Inputs: $\vee$-list of nodes, $\varepsilon$-list of edges, $s$-source node, $Y$-observation matrix,Returns: $\pi$[$v$]-predecessor of each node, $w$-edge weight matrix}
    \State $P$ := NIL
    \State $Q$ := queue($\vee$)
    \State $k$ := 0
   \State $p\varepsilon$[$s$] = 0
    \State $w$[$p\varepsilon$[$s$], $s$] = 0
    %\pagebreak
    $initialise\_single\_source$($\vee,\varepsilon,s$)
    \While{$Q != $0}
       $u$ := Extract min-priority queue($Q$)
       $p\vee$[$u$], $npred$ = $find\_prev$($u,s$)
       $k$ = $npred$+1
       $pW$ = $w$[$p\varepsilon$[$u$], $u$]
       $P$ := $P$ $\bigcup u$
       \For{each $v \in Adj$[$u$]}
         \State $w$[$u$,$v$] = $find\_cost$($u$,$v$,$k$, $pW$,$Y$($k$))
         \State $relax$($u$,$v$,$w$)
         \EndFor
       \EndWhile
    \EndFunction
 \end{algorithmic}
%\end{multicols}
\end{algorithm}
\section{Discussion and Conclusion}
\label{last}
The travel times of edges are identified as one of the cost coefficients in internal automated logistics. A formulation is devised to estimate travel times online during path computation considering its time-varying components. Moreover, suitable observations for travel time are recorded in scenarios with analogy to real factory in a scaled platform developed in the laboratory. They are instrumental for feeding into estimation algorithms to estimate travel time. Path is found using Dijkstra's algorithm based on both heuristic weights of edges and estimated travel times of edges as weights. Results show that paths computed using travel time as weights of edges have lesser total path cost than that of obtained by heuristic weights. 

It is interesting to mention here that route planning is done in Tesla's new X 75D model cars according to battery need. The route planner proposes breaks of variable times to enable recharging while travellers enjoy recess in driving. Thus, states of battery and environment are considered for optimal route planning in these models. This example signifies the current work in enabling the cost parameter to implicitly indicate the battery charge and environmental factors.

In this work, the cost of traversing every edge is estimated, which facilitates to apply deterministic path planning algorithms like Dijkstra's algorithm, Bellmont-Ford algorithm \textit{et~cetera}. Many industries (like BlueBotics \cite{Blue:2009}) use topology maps to describe the floor and employs a depth-first search to generates a length-optimal path using deterministic path planning algorithms. This work is complementary to this approach where the travel times can be used as path determining factor in those deterministic algorithms without changing any model of computation or architecture.

The approach used in single-task case in this work can be extended in multi-task scenarios for a MR, where cost coefficient for different tasks has to be found out. This is a direction for future consideration and it could be extended to every MR in the system. During the run-time of MRS, every estimated value of travel time has context depending on various environmental and inherent factors. Travel time of one MR can provide contextual information to other MRs in an multi-robot system (MRS) and contribute in estimating travel time for them. This enhances further investigation towards implementing collaborative or collective intelligence in MRS to have cost efficient coordination of the MRS. 
\ifCLASSOPTIONcaptionsoff
  \newpage
\fi
\bibliographystyle{plain}
\bibliography{DR_Jrnl_IEEE.bib}
\end{document}